\documentclass{cimento}

\setcopyright{FNAL on behalf of the CDF and D0 Collaboration}

\usepackage{graphicx}  
\usepackage{xspace}

\newcommand{\MET}{\mbox{$\raisebox{.3ex}{$\not\!$}E_T$}}

\newcommand{\ttbar}     {\mbox{$t\bar{t}$}\xspace}

\newcommand{\ppbar}     {\mbox{$p\bar{p}$}\xspace}
\newcommand{\ljets}     {\mbox{$\ell$+jets}\xspace}

\newcommand{\singletop} {\sc singletop}

\newcommand{\pythia}    {\mbox{\textsc{pythia}}}

\newcommand{\geant}     {{\sc{geant}}}
\newcommand{\alpgen}    {\mbox{\textsc{alpgen}}}
\newcommand{\mcfm}      {\sc mcfm}

\newcommand{\xsectev}{1.29}
\newcommand{\xsecteverrorup}{+0.26}
\newcommand{\xsecteverrordown}{-0.24}

\title{Observation of the $s$ channel and other studies of single top quarks at the Tevatron}

\author{ R.~C.~Group}

\instlist{\inst{Fermilab - Batavia, IL, USA ; University of Virginia - Charlottesville, VA, USA; \\ On behalf of the CDF and D0 Collaborations}}

\PACSes{
\PACSit{14.65.Ha}{13.85.Qk}
}

\begin{document}

\maketitle

\begin{abstract}
First observation of single-top-quark production in the $s$ channel is
reported.  The result is based on the combination of the CDF and D0
measurements of the cross section in proton-antiproton collisions at a
center-of-mass energy of 1.96~TeV.  A summary of other recent
single-top-quark results are also included.
\end{abstract}

\section{Introduction}

The top quark, with a mass of
$m_t=173.2\pm0.9$~GeV~\cite{ref:topmassTeV}, is the most massive of the elementary particles of the standard model
(SM).  In proton-antiproton collisions, top quarks can be produced singly through electroweak
interactions and this process provides a unique opportunity to test the standard model and search for non-SM physics. 

In the SM, the single-top-quark production cross section is predicted to be
proportional to the square of the magnitude of the quark-mixing
Cabibbo-Kobayashi-Maskawa matrix~\cite{ref:ckm} element $V_{tb}$.  Consequently, measurements of the single-top-quark production rate could be sensitive to contributions from a fourth generation of quarks~\cite{ref:FourthGen1,ref:FourthGen2}.

\begin{figure}[!h!tbp]
\begin{center}
\includegraphics[width=0.48\textwidth]{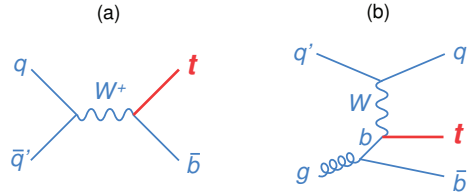}
\caption{Dominant Feynman diagrams for (a) $s$-channel
  and (b) $t$-channel 
single-top-quark production at the Tevatron.}
\label{fig:feynman_diagrams}
\end{center}
\end{figure}

The two dominant production modes of the single-top-quark process are shown in Fig.~\ref{fig:feynman_diagrams} and are sensitive to different classes of SM extensions~\cite{ref:Tait:2000sh}. The $s$-channel process~\cite{ref:singletop-cortese,ref:schannel-kidonakis}, in which an
intermediate $W$ boson decays into a top (antitop) quark and an antibottom (bottom) quark,
is sensitive to contributions from additional heavy bosons~\cite{ref:WPrime1,ref:WPrime2,ref:WPrime_CDF,ref:WPrime_D0}; the
$t$-channel process~\cite{ref:singletop-willenbrock,ref:singletop-yuan,ref:tchannel-kidonakis}, in which a bottom quark transforms into a top quark by exchanging a
$W$ boson with another quark, is more sensitive to flavor-changing neutral
currents~\cite{ref:FCNC1,ref:FCNC2,ref:FCNC_CDF,ref:FCNC_D0}. Independently studying the production rate of these
channels provides more restrictive constraints on SM extensions than just studying the
combined production rate.

  Single-top-quark production was first observed independently by the
CDF and D0 experiments in
Refs.~\cite{ref:stop-obs-2009-cdf,ref:cdf-prd-2010}
and~\cite{ref:stop-obs-2009-d0,ref:stop-2011-d0}, respectively. These
results were based on an inclusive search targeting the combined contribution of $s$-channel and
$t$-channel production.

Several differences in $s$- and $t$-channel event properties can be
used to distinguish them from one another.  For example, events
originating from $t$-channel production typically contain one
light-flavor jet in the forward detector region (at large
pseudorapidity).  The direction of this jet (whether it is along the
proton or anti-proton direction) is also correlated with the charge of
the lepton.  These properties are very powerful for distinguishing them
from events associated with $s$-channel production and other SM
background processes.  Moreover, events from the $s$-channel process
are more likely to contain two jets which can be identified as
originating from $b$ quarks ($b$ jets). Hence, single-top-like events
with two identified $b$ jets are more likely to have originated from
$s$-channel production.

 The $t$-channel single-top quark production was first observed in
 2011 by the D0 experiment~\cite{ref:t-channel-new}, and later
 confirmed in 2012 by the ATLAS and CMS experiments at the
 LHC~\cite{ref:atlas-tchannel,ref:cms-tchannel}.  The $s$-channel
 process has not yet been observed.  Because of the smaller production
 cross section and larger backgrounds, it is more difficult to isolate
 it compared to the $t$-channel process in proton-antiproton
 collisions. It is even more difficult at the LHC, as proton-proton
 collisions yield a significantly smaller signal-to-background ratio
 compared to the Tevatron. In fact, to date the LHC experiments have
 only reported unpublished upper limits on the cross section.

\section{Single Top Event Selection and Background Model}
  Since the magnitude of the $W$-top-bottom quark coupling is much
larger than the $W$-top-down and $W$-top-strange quark
couplings, each top quark is assumed to decay exclusively into a
$W$~boson and a $b$ quark. We seek events in which the $W$ boson decays leptonically ($W
\to \ell \nu_\ell$).  Online event selection is based on identifying an isolated high-$p_T$ lepton or a large imbalance of transverse energy \MET.  The offline event selection is split into two distinct topologies, both designed to select single-top-quark events in which the
$W$~boson decays leptonically.

  One final-state topology (\ljets), analyzed by both the CDF and D0
  collaborations, contains single-top-quark events in which the
  charged lepton from the $W$~boson decay is identified.  Events are
  selected with an identified electron or muon, two or more jets, and
  \MET.  At least one of the jets must be identified as a $b$ jet~\cite{ref:CDFbtag,ref:D0btag}.
  Additional selection criteria are applied to exclude kinematic
  regions that are difficult to model, and to minimize the quantum
  chromodynamics (QCD) multijet background.

The other final-state topology, analyzed by the CDF collaboration,
involves~\MET\ and jets, the event selection is similar to \ljets, but
no reconstructed isolated charged leptons are allowed in the
event(\MET+jets).  This additional sample increases the acceptance for
$s$-channel signal events by encompassing those in which the $W$-boson
decay produces a muon or electron that is either not reconstructed or
not isolated, or a hadronically decaying tau lepton that is
reconstructed as a third jet. In order to reduce the large multijet background in this channel, a neural-network
event selection is optimized to preferentially select signal-like
events. 

Events passing the \ljets and \MET+jets selections are further
separated into independent analysis channels based on the number of
reconstructed jets as well as the number and quality of $b$-tagged
jets.  Each of the analyzed channels has a different
background composition and signal ($s$) to
background ($b$) ratio. Analyzing them separately enhances the
sensitivity to single-top-quark 
production.

 Both collaborations use Monte Carlo~(MC) generators to simulate the
kinematic properties of signal and background events, except in the
case of multijet production, for which the model is derived from data.
 Single-top-quark signal events are modeled with 
next-to-leading-order (NLO) accuracy in the strong coupling constant
$\alpha _s$~\cite{ref:POWHEG2009,ref:singletop-mcgen}.

Kinematic properties of background events associated with the $W$+jets
and $Z$+jets processes 
are simulated using a leading-order MC
generator~\cite{ref:alpgen}, and those of diboson processes ($WW$, $WZ$
and $ZZ$) and {\ttbar} are modeled using MC~\cite{ref:pythia}.   In all cases {\pythia} is used
to model proton remnants and simulate the
hadronization of all generated partons.  All MC events are
processed through {\geant}-based detector simulations~\cite{ref:geant}.

Predictions for the normalization of simulated background-process
contributions are estimated using both simulation and data.
Data are used to normalize the $W$ plus
light-flavor and heavy-flavor jet contributions using enriched
$W$+jets data samples
that have negligible signal.   All other
simulated background samples are normalized to the theoretical cross
sections at NLO combined with next-to-next-to-leading log (NNLL)
resummation~\cite{ref:tchannel-kidonakis} for $t$-channel single-top-quark
production, at next-to-NLO~\cite{ref:ttbar-xsec} for {\ttbar}, at
NLO~\cite{ref:mcfm} for $Z$+jets and diboson production, and 
including all relevant higher-order QCD and electroweak corrections
for Higgs-boson production~\cite{ref:higgs-xsec}.
Differences observed between simulated events and data in lepton and
jet reconstruction efficiencies, resolutions, jet-energy scale (JES), and
$b$-tagging efficiencies are adjusted in the simulation to match the
data, through correction functions obtained from measurements in
independent data samples.

\section{Analysis Overview}

D0 has optimized the single top analysis from Ref.~\cite{ref:t-channel-new} to enhance the sensitivity to the $s$-channel in the $\ell\nu bb$ final state, and has improved the selection by employing a newer more efficient $b$-tagging algorithm. In the new analysis~\cite{ref:Abazov:2013qka} the events are split into four independent channels depending on the jet multiplicity and the number of $b$-tagged jets. Three multivariate methods are applied to the selected data: Boosted Decision Trees (BDT), Bayesian Neural Networks (BNN), and Matrix Element probability calculations (ME). They are optimized to measure the $s$ and the $t$ channels separately in all four analysis channels.  The BDT method uses up to 30 kinematic variables, the BNN uses the objects' four momenta with a few other variables such as the $b$-tag output of jets, $q_
{\ell} \times \eta$(light jet), and the invariant $W$~boson mass, and the ME uses the objects' four vectors. The three methods are not fully correlated ($\sim 75\%$) and their discriminant outputs are therefore combined by a second BNN which gives a better expected sensitivity than any individual method. One of the most important improvements in this analysis is the development of a new method to measure the $s$- and the $t$-channels independently, without assuming the SM $s/t$ ratio when calculating the $s+t$ cross section.  This flexible analysis is used for all of the D0 results discussed below.   

The CDF results discussed below are based on four different analyses. In an analysis from a few years ago~\cite{ref:CDFnote75}, a NN is used to make single top quark measurements in the $\ell+jets$ sample.  Events with two or three jets and one or two $b$ tags are considered. In the category with two tags and two $b$ jets, the NN is optimized for $s$ channel, while in the other three categories the NN is optimized for $t$ channel events. More recently, an analysis has been optimized for $s$ channel in all categories~\cite{ref:Aaltonen:2014qja} of the lepton plus two jet sample.   The analysis is based on the SM Higgs boson search~\cite{ref:cdf_WH}, which shares the same final state: $\ppbar \to WH \to \ell\nu b\bar{b}$. There are three lepton categories: central tight electrons, central tight muons, and extended muons (loose muons and isolated track lepton-candidates). The identification of $b$ jets has also improved with respect to previous searches~\cite{ref:CDFbtag}. The selection requires exactly two jets, at least one of which is required to be $b$-tagged. The analysis is then split in four channels according to the different $b$-tagged properties of the jets. One operational point (denoted T for tight) has 42\% $b$-jet efficiency and 0.9\% mistag rate,
and a second operational point (denoted L for loose) has 70\% $b$-jet efficiency and 9\% misidentification rate per jet. Events are therefore split according to the four orthogonal categories: TT, TL, T, and LL. This improves the sensitivity of the analysis by separating regions of the phase space with different signal-to-background ratios. CDF also has two analyses based on the \MET+jets topology: one for the combined $s+t$ analysis~\cite{ref:CDFnoteMETjet}, and one that is optimized for the $s$ channel~\cite{ref:Aaltonen:2014xta}.

\section{Inclusive Single Top Production Studies}

\begin{figure}[!h!tbp]
\begin{center}
\includegraphics[width=0.68\textwidth]{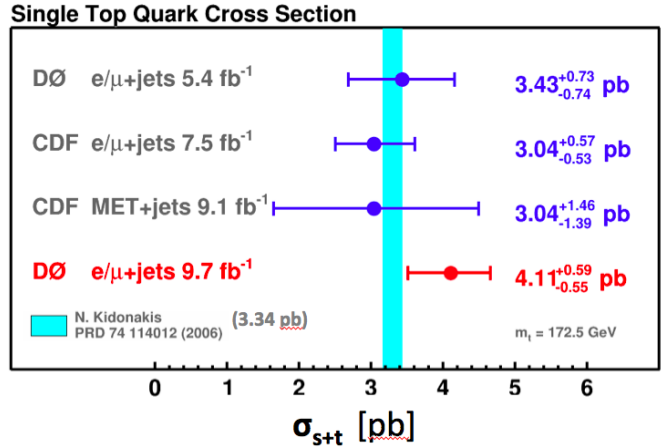}
\caption{Summary of recent measurements of the combined single top cross section measurements at the Tevatron ($s$+$t$ channel).}
\label{fig:s_plus_t_summary}
\end{center}
\end{figure}

 For each analysis multivariate discriminants are trained to separate the single top signal from the background processes.  A binned likelihood technique using the multivariate discriminant shapes is employed in order to extract the most likely value for the cross section measurement.  The most recent measurements of the combined $s$-channel plus $t$-channel cross section measurements from the Tevatron are summarized in Fig.~\ref{fig:s_plus_t_summary}.

\section{Constraints on the CKM Element $V_{tb}$}

  From the combined $s+t$ cross section, D0 calculates the value
 of $|V_{tb}|$ without any assumption on the number of generations or the unitarity of the CKM matrix, and the result is: $|V_{tb}| = 1.12^{+0.09}_{-0.08}$, pr $|V_{tb}| > 0.92$ at $95\%$~C.L., assuming a flat prior within $0 \leq |V_{tb}|^2 \leq 1$. This is currently the most stringent lower limit on $|V_{tb}|$ from the single top measurements at the Tevatron.  

A past CDF and D0 combination measured $|V_{tb}| = 0.88^{+0.07}_{-0.07}$~\cite{ref:Group:2009qk}.  A combination is also planned for the updated analyses.  

\section{Two-dimensional $s$- and $t$-channel Measurements}

\begin{figure}[!h!tbp]
\begin{center}
\includegraphics[width=0.68\textwidth]{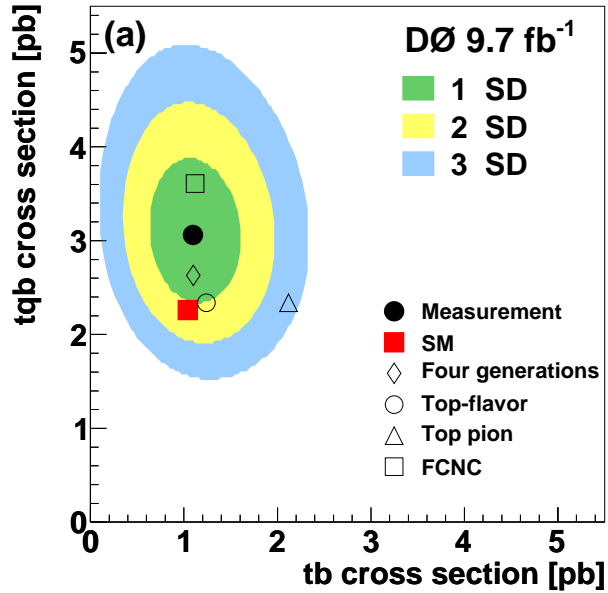}
\caption{The D0 two-dimensional posterior probability density as a function of the $s$-channel ($tb$) and the $t$-channel ($tqb$) cross sections. Several models beyond the SM are shown for reference~\cite{ref:FourthGen2,ref:Tait:2000sh}. }
\label{fig:2D_D0}
\end{center}
\end{figure}

In addition to the inclusive cross section and the $V_{tb}$ extraction, both CDF and D0 perform a two-dimensional fit for the $s$ and $t$ channel cross sections.  As an example, the two-dimensional posterior probability density as a function of the $s$-channel and the $t$-channel cross sections from the D0 analysis is shown in Figure~\ref{fig:2D_D0}. The predictions from several exotic models are shown on the figure.  CDF and D0 are planning to produce a combined 2-D fit employing the full data set of both experiments.

\section{Analyses Optimized for $s$-channel Significance}

\begin{figure}[!h!tbp]
\begin{center}
\includegraphics[width=0.68\textwidth]{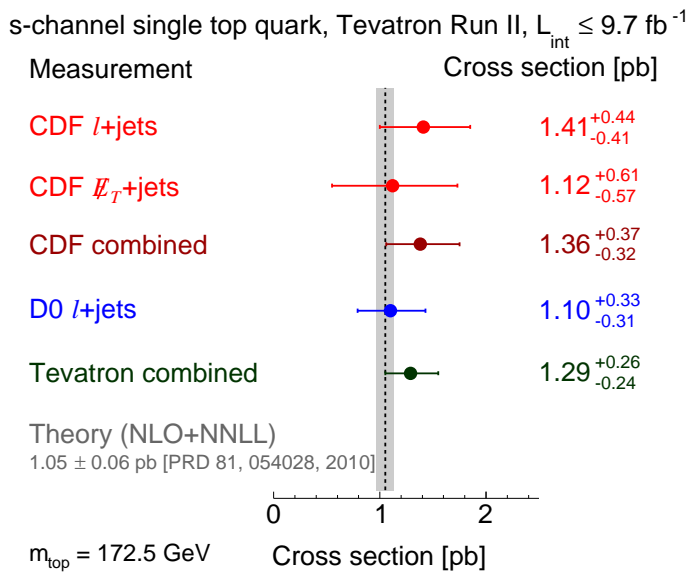}
\caption{Summary of recent measurements of the $s$-channel single top cross section measurements and the Tevatron combination.}
\label{fig:s_summary}
\end{center}
\end{figure}

  In the summer of 2013, D0 reported the first evidence for the the $s$-channel process with a significance greater than 3~SD: the expected and observed significances are 3.7~SD~\cite{ref:Abazov:2013qka}. Later that summer, CDF also reported evidence for the $s$ channel with an observed significance of 3.8~SD~\cite{ref:Aaltonen:2014qja}.  Recently, the collaborations have combined their measurements~\cite{ref:CDF:2014uma}.  Figure~\ref{fig:s_summary} summarizes the most recent $s$-channel cross section measurements from each experiment and the Tevatron combination.  The combined value of the $s$-channel single-top-quark production
cross section 
is measured to be $\sigma_s = \xsectev^{\xsecteverrorup}_{\xsecteverrordown}$~pb, in agreement with
the SM expectation of $\sigma_ s^{SM} = 1.05 \pm
0.06$~pb ($m_t=172.5$~GeV)~\cite{ref:schannel-kidonakis}.

\begin{figure}[!h!tbp]
\begin{center}
\includegraphics[width=0.68\textwidth]{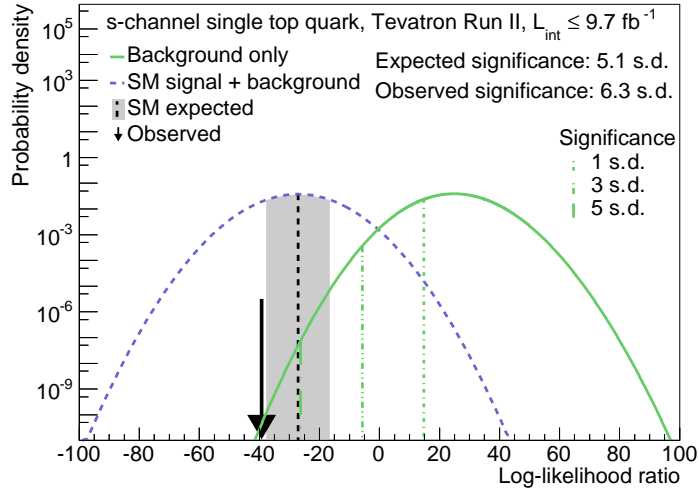}
\caption{
Log-likelihood ratios for the background-only
  (solid green line) and SM-signal-plus-background (dashed blue)
  hypotheses from the combined measurement.  The significance of the measurement is well over 5 SD and represents the first observations of $s$-channel single top quark production.}
\label{fig:pValue}
\end{center}
\end{figure}

 The statistical significance of the combined result was quantified
  by an asymptotic log-likelihood ratio approach
  (LLR)~\cite{ref:loglhood}, including systematic uncertainties (see Fig.~\ref{fig:pValue}).  The
  probability to measure an $s$-channel cross section of at least the
  observed value in the absence of signal is $1.8 \times 10^{-10}$,
  corresponding to a significance of 6.3 standard deviations, well
  beyond the standard to claim discovery.

\section{Summary and Conclusions}

  Single-top-quark production was first observed at the Tevatron in 2009 independently by the CDF and D0 experiments. Soon after, the D0 collaboration reported a significance of greater than 5 standard deviations for the $t$-channel process.  Both experiments have produced constraints on $V_{tb}$.  These constraints are still competitive with those from the LHC, and an effort to combine the CDF and D0 constrains is underway.       

   The LHC experiments have confirmed the $t$-channel process, but measurements of the $s$-channel process are much more difficult in the LHC environment than at the Tevatron.  These proceedings document the first conference announcement (La Thuile 2014) that the CDF and D0 collaborations combined their datasets and observed the $s$-channel process with a significance of over 6 standard deviations.

\acknowledgments

\end{document}